\documentclass{ws-ijmpe}
\usepackage[super,compress]{cite}
\usepackage{subcaption}  
\usepackage{comment} 
\usepackage{epstopdf}  %
\usepackage{float}  
\usepackage{epsfig}
\usepackage{amssymb}
\usepackage{float}
\usepackage{color}
\usepackage{siunitx}
\usepackage{graphicx}
\usepackage{hyperref}
\hypersetup{
	colorlinks   = true, 
	urlcolor     = blue, 
	linkcolor    = blue, 
	citecolor   = blue 
}
\usepackage{makecell}
\begin{document}

\markboth{Y. EL BASSEM $\&$ M. OULNE}{Ground state properties and shape evolution in Pt isotopes....}

\catchline{}{}{}{}{}

\title{Ground state properties and shape evolution in Pt isotopes within the covariant density functional theory}

\author{Y. EL BASSEM\footnote{corresponding author.} ~and M. OULNE$^\dagger$}

\address{High Energy Physics and Astrophysics Laboratory, Department of Physics, \\Faculty of Sciences SEMLALIA, Cadi Ayyad University,  \\P.O.B. 2390, Marrakesh, Morocco.\\
$^*$younes.elbassem@ced.uca.ma\\
$^\dagger$oulne@uca.ma}

\maketitle

\begin{history}
\received{Day Month Year}
\revised{Day Month Year}
\end{history}

\begin{abstract}
	In this work, the ground-state properties of the platinum isotopic chain, $^{160-238}$Pt are studied within the covariant density functional theory. The calculations are carried out for a large number of even-even Pt isotopes by using the density-dependent point-coupling and the density dependent meson-exchange effective interactions. All ground-state properties such as the binding energy, separation energy, two-neutron shell gap, rms-radii for neutrons and protons and quadrupole deformation are discussed and compared with available experimental data, and with the predictions of some nuclear models such as the Relativistic Mean Field (RMF) model with NL3 functional and the Hartree Fock Bogoliubov (HFB) method with SLy4 Skyrme force. The shape phase transition for Pt isotopic chain is also studied. Its corresponding total energy curves as well as the potential energy surfaces confirm the transition  from prolate to oblate shapes at $^{188}$Pt contrary to some studies predictions and in agreement with others. Overall, a good agreement is found between the calculated and experimental results wherever available.
\end{abstract}


\ccode{PACS numbers: 21.10.−k, 21.10.Dr, 21.10.Ft, 21.60.−n}


\section{Introduction}

One of the main  aims of research in nuclear physics is to make reliable predictions for the ground state nuclear properties of all nuclei in the periodic table with one nuclear model. For this purpose, several approaches have been developed and they can be classified into three categories: The first one called
the macroscopic models such as the Bethe-Weizsäcker mass formula~\cite{samanta2002}. The second one is the macroscopic-microscopic models such as the Finite Range Droplet Model
(FRDM)~\cite{moller1995}. The last one consists of the microscopic models such as the conventional Hartree Fock method~\cite{skyrme1956,decharg1980,bassem2015,bassem2016,antonov2017} with effective density-dependent interactions and its relativistic analog, the relativistic mean field (RMF) theory~\cite{walecka1974,reinhard1989}; the former is based on the nonrelativistic kinematics, in which the density-dependent and the spin-orbit interactions are important ingredients, while the latter is based on the relativistic kinematics, 
where the ingredients are the nucleons, the mesons and their interactions.
Therefore, the spin appears automatically and the interplay of the scalar and the vector potentials leads naturally to a proper spin orbit interaction and appropriate shell structure. The RMF theory  has received wide attention due to its successful description of lots of properties of nuclei not only in the valley of stability but also far away from it.~\cite{meng1998,zhou2003,meng1999,geng2004,ring1996,meng1996,ginocchio1997}.

Density functional theories (DFT’s) are extremely useful in understanding nuclear many-body dynamics. Among different nuclear DFTs, the covariant density functional theory (CDFT)~\cite{niki2014,lalazissis2005,rocamaza2011,niki2002} based on the energy density functionals (EDFs) is one of the most attractive and is very successful in  describing very well the ground and excited states throughout the
chart of nuclei~\cite{abusara2012,agbemava2015,agbemava2017} as well as in the nuclear structure analysis~\cite{meng2015,matev2007,afanasjev2008}.
In Ref.~\refcite{Agbemava2014}, the global performance of some covariant
energy density functionals on some nuclear observables was analyzed.

In this work, we are interested in calculating and analyzing some ground-state properties of even-even Pt isotopes, N=82-160, within the framework of the covariant density functional theory by using two of the state-of-the-art functionals which provide a complete and an accurate description of different ground states and excited states over the whole nucleic chart~\cite{karim2015,afanasjev2010,elbassem2019}, namely : The density-dependent point-coupling DD-PC1~\cite{niki2008} and the density-dependent meson-exchange DD-ME2~\cite{DD-ME2}.

The paper is organized in the following way: The covariant density functional theory and details of the numerical calculations are presented in section~\ref{Theoretical Framework}. 
Section~\ref{Results and Discussion} is devoted to present our results and discussion.
Finally, the conclusions of this study are presented in section~\ref{Conclusion}.

\section{Covariant density functional theory}
\label{Theoretical Framework}
Two classes of covariant density functional models are used throughout this paper: the density-dependent point-coupling (DD-PC) model and the density-dependent meson-exchange (DD-ME) model. 
The first has a finite interaction range and has been fitted 
to binding energies and radii of spherical nuclei; while the latter uses a zero-range interaction and has been fitted to nuclear matter data and for finite nuclei only to binding energies  of a large range of deformed nuclei.\\

In the meson-exchange model, the nucleus is considered as a system of Dirac nucleons which interact via the exchange of mesons with finite masses leading to finite-range interactions~\cite{typel1999,lalazissis2009}.
The standard Lagrangian density with medium dependence vertices that defines the meson-exchange model~\cite{gambhir1990} is given by:
\begin{eqnarray}
\mathcal{L}  &  =\bar{\psi}\left[
\gamma(i\partial-g_{\omega}\omega-g_{\rho
}\vec{\rho}\,\vec{\tau}-eA)-m-g_{\sigma}\sigma\right]  \psi\nonumber  +\frac{1}{2}(\partial\sigma)^{2}-\frac{1}{2}m_{\sigma}^{2}\sigma^{2}\\
&
-\frac{1}{4}\Omega_{\mu\nu}\Omega^{\mu\nu}+\frac{1}{2}m_{\omega}^{2}\omega
^{2}\label{lagrangian}  -\frac{1}{4}{\vec{R}}_{\mu\nu}{\vec{R}}^{\mu\nu}+\frac{1}{2}m_{\rho}%
^{2}\vec{\rho}^{\,2}-\frac{1}{4}F_{\mu\nu}F^{\mu\nu}
\end{eqnarray}
where $m$ is the bare nucleon mass and $\psi$ denotes the Dirac spinors. $m_\rho$, $m_\sigma$ and $m_\omega$ are the masses of  $\rho$ meson,  $\sigma$ meson and  $\omega$ meson, 
with the corresponding coupling constants for the mesons to the nucleons as $g_\rho$, $g_\sigma$ and $g_\omega$ respectively, and $e$ is the charge of the proton.\\

The point-coupling model represents  an alternative formulation of the self-consistent RMF framework~\cite{manakos1988,rusnak1997,brvenich2002,zhao2010}. The Lagrangian for the DD-PC model~\cite{niki2008,nikolaus1992} is given by:
\begin{eqnarray}
&\mathcal{L}    =\bar{\psi}\left(i\gamma \cdot \partial-m\right)  \psi
-\frac{1}{2}\alpha_S(\hat\rho)\left(\bar{\psi}\psi\right)\left(\bar{\psi}\psi\right)-\frac{1}{2}\alpha_V(\hat\rho)\left(\bar{\psi}\gamma^{\mu}\psi\right)\left(\bar{\psi}\gamma_{\mu}\psi\right)\label{Lag-pc}\\
&
-\frac{1}{2}\alpha_{TV}(\hat\rho)\left(\bar{\psi}\vec\tau\gamma^{\mu}\psi\right)\left(\bar{\psi}\vec\tau\gamma_{\mu}\psi\right)-\frac{1}{2}\delta_S\left(\partial_v\bar{\psi}\psi\right)\left(\partial^v\bar{\psi}\psi\right) - e\bar\psi\gamma \cdot A
\frac{(1 - \tau_3)}{2}\psi .\nonumber
\end{eqnarray}
The Eq.~(\ref{Lag-pc}) contains the free-nucleon Lagrangian, the point coupling interaction terms, and the coupling of the proton to the electromagnetic field, the derivative terms account for the leading effects of finite-range interaction
which are important in nuclei. 

In the present work, we have used the recently developed density-dependent meson-exchange relativistic energy functional DD-ME2~\cite{DD-ME2} and the very successful density-dependent point-coupling interaction DD-PC1~\cite{niki2008} with the parameter sets listed in Tables~\ref{tab1}~and~\ref{tab2} respectively.

\begin{table}[!ht]
	\caption{ The parameters of the  DD-ME2 functional. 
		\label{tab1}}
	\begin{center}%
		\begin{tabular}
			[c]		{@{\hspace{0pt}}c@{\hspace{24pt}}@{\hspace{24pt}}c@{\hspace{0pt}}}
			\hline Parameter &  DD-ME2 \\\hline
			$m$   & 939  \\
			$m_{\sigma}$  & 550.1238 \\
			$m_{\omega}$  & 783.000  \\
			$m_{\rho}$    & 763.000 \\
			$g_{\sigma}$  &  10.5396\\
			$g_{\omega}$  & 13.0189\\
			$g_{\rho}$    & 3.6836  \\
			$a_{\sigma}$  & 1.3881 \\
			$b_{\sigma}$  & 1.0943 \\
			$c_{\sigma}$  & 1.7057 \\
			$d_{\sigma}$  & 0.4421 \\
			$a_{\omega}$  & 1.3892 \\
			$b_{\omega}$  & 0.9240 \\
			$c_{\omega}$  & 1.4620 \\
			$d_{\omega}$  & 0.4775 \\
			$a_{\rho}$    & 0.5647 \\
			\hline
		\end{tabular}
	\end{center}
\end{table}

\begin{table}[!ht]
	\caption{The parameters of the DD-PC1 functional.}%
	\label{tab2}
	\begin{center}
		\begin{tabular}
			[c]		{@{\hspace{0pt}}c@{\hspace{24pt}}@{\hspace{24pt}}c@{\hspace{0pt}}}
			\hline Parameter &  DD-PC1 \\\hline
			$m$ &   939\\
			$a_{\sigma}$ & -10.04616\\
			$b_{\sigma}$ & -9.15042\\
			$c_{\sigma}$ & -6.42729\\
			$d_{\sigma}$ &  1.37235\\
			$a_{\omega}$ &  5.91946\\
			$b_{\omega}$ &  8.86370\\
			$d_{\omega}$ &  0.65835\\
			$b_{\rho}$   &  1.83595\\
			$d_{\rho}$   &  0.64025\\
			\hline
		\end{tabular}
	\end{center}
\end{table}

\section{Results and Discussion}
\label{Results and Discussion}

In this section, we present the ground-state properties of $^{160-238}$Pt nuclei obtained in the framework of the CDFT by using the interactions DD-ME2~\cite{DD-ME2} and DD-PC1~\cite{niki2008}. 
Our results are compared with the available experimental data, the predictions of the RMF model with NL3~\cite{lalazissis1999} functional and with the results of HFB theory with SLy4 Skyrme force calculated by using the computer code HFBTHO v2.00d~\cite{Stoitsov2013} as in Ref~\refcite{bassem2017}. 

\subsection{Binding energy}

\begin{figure}[ht]
	\centering \includegraphics[scale=0.5]{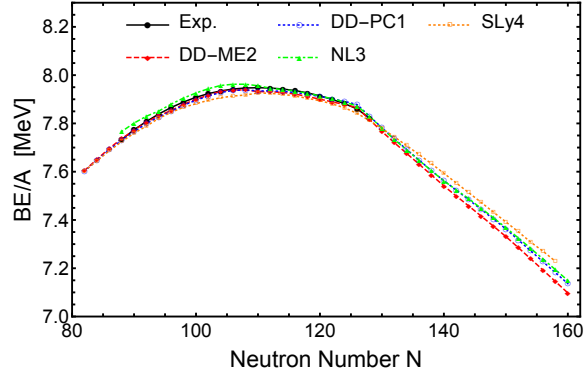}
	\caption{(Color online) The binding energies per nucleon for even-even $^{166-238}$Pt isotopes.}
	\label{BEexp}
\end{figure}

Binding energy (BE) is directly related to the stability of nuclei and is a very important quantity in nuclear physics. 
The binding energies per nucleon (BE/A) of ground states for platinum isotopes, $^{160-238}$Pt, are presented in Fig.~\ref{BEexp} as a function of the neutron number N. The available
experimental data~\cite{wang2012} as well as the predictions of the RMF(NL3)~\cite{lalazissis1999} and HFB(SLy4) theories are also shown for comparison.

It can be clearly seen from Fig. \ref{BEexp} that the theoretical predictions reproduce the experimental data accurately and, qualitatively, all curves show a similar behavior.

\begin{figure}[ht]
	\centering \includegraphics[scale=0.5]{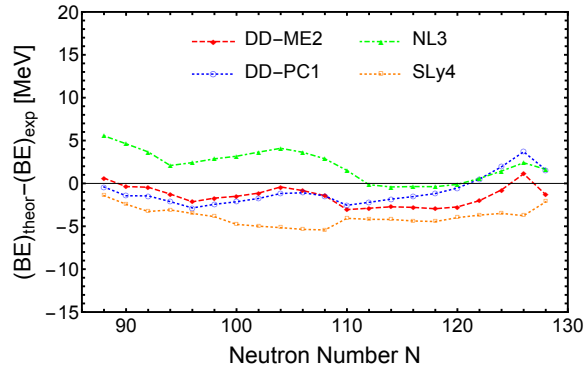}
	\caption{(Color online) Differences between theoretical total binding energies and experimental values for even-even Pt isotopes.}
	\label{Delta_BE}
\end{figure}

In order to provide a further check of the accuracy of our results, the differences between the experimental total BE for even-even platinum isotopes and the calculated results obtained in this work are presented in Fig.~\ref{Delta_BE} as a function of the neutron number N. 

The comparison of the available experimental total binding energies of the ground state for $^{166-238}$Pt with the present calculations and with the other theoretical models, is also done by the root mean square (rms) deviation tabulated in Table~\ref{rms}.

\begin{table}[!htb]
	\caption{The rms deviations of the total binding energies of platinum isotopes.\label{rms}}
	\centering
	\begin{tabular}
		{@{\hspace{0pt}}c@{\hspace{24pt}}c@{\hspace{24pt}}c@{\hspace{0pt}}@{\hspace{24pt}}c@{\hspace{0pt}}}
		\hline\noalign{\smallskip}
		DD-ME2 	&   DD-PC1	&  NL3		 & SLy4\\ 
		\noalign{\smallskip}\hline\noalign{\smallskip}
		1.864 	&	1.880	&	2.755	 & 	4.019	\\  
		\noalign{\smallskip}\hline
	\end{tabular}
\end{table}

As we can see from Fig.~\ref{Delta_BE} and Table~\ref{rms}, DD-ME2 exceeds in accuracy the other nuclear models DD-PC1, NL3 and SLy4.

\subsection{Two neutron separation energy ($S_{2n}$) and shell gap ($\delta_{2n}$)}

In the present work, we have calculated two-neutron separation energies, 
$S_{2n}(N,Z)$ = BE$(N,Z)$ - BE$(N - 2,Z)$, for Pt isotopes by using the density-dependent effective interactions DD-PC1 and DD-ME2. 

\begin{figure}[ht]
	\centering \includegraphics[scale=0.52]{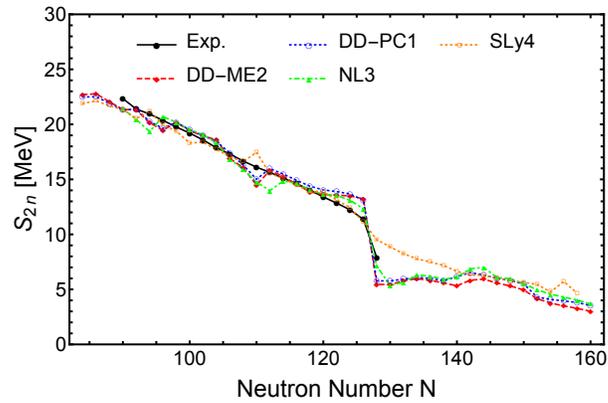}
	\caption{(Color online) The two-neutron separation energies, S$_{2n}$, for Pt
		isotopes.}
	\label{S2n}
\end{figure}

In Fig.~\ref{S2n}, we display the calculated S$_{2n}$ of  even-even platinum isotopes, as a function of the neutron number N, in comparison with the available experimental data~\cite{wang2012} and the predictions of RMF(NL3)~\cite{lalazissis1999} and HFB(SLy4).

As one can see from Fig.~\ref{S2n}, the results of the two density-dependent models DD-ME2 and DD-PC1 as well as those of NL3 and SLy4 reproduce the experimental data quite well except some small discrepancies which are mainly due to the missing beyond mean field corrections~\cite{Litvinova2006}. S$_{2n}$ gradually decreases with N, and a sharp drop is distinctly seen at N = 126 in both experimental and theoretical curves, except in SLy4 curve, which corresponds to the closed shell at this magic neutron number. 
The discontinuity of S$_{2n}$ at this nucleus indicates that it corresponds to a critical point of first order phase transition to prolate shape.

\begin{figure}[!ht]
	\centering \includegraphics[scale=0.55]{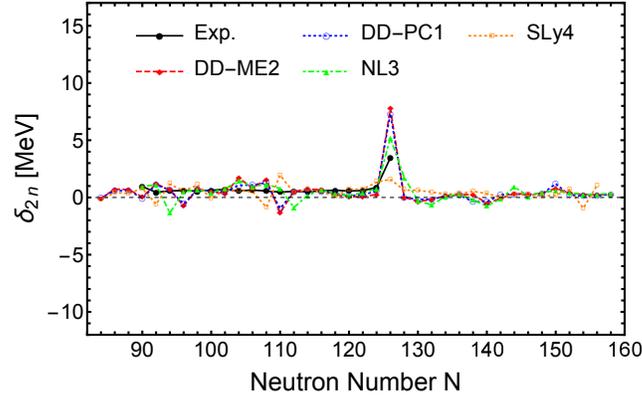}
	\caption{(Color online) The two-neutron shell gap $\delta_{2n}$ for even-even $^{94-168}$Pt isotopes.}
	\label{ShellGap}
\end{figure}

A more sensitive observable for locating the shell closure is the two-neutron shell gap $\delta_{2n}=S_{2n}(N,Z)-S_{2n}(N+2,Z)$, also known as the S$_{2n}$~differential. 
$\delta_{2n}$ is shown in Fig.~\ref{ShellGap} as a function of the neutron number N. The strong peaking in the two-neutron shell gap ($\delta_{2n}$) clearly seen at N~=~126 further supports the shell closure at this neutron magic number as shown in Fig.~\ref{S2n} by two-neutron separation energy ($S_{2n}$). Here again, only the SLy4 curve does not show a strong peak at the neutron magic number N~=~126.

\subsection{Quadrupole deformation} 

\begin{figure}[!htb]
	\centering \includegraphics[scale=0.6]{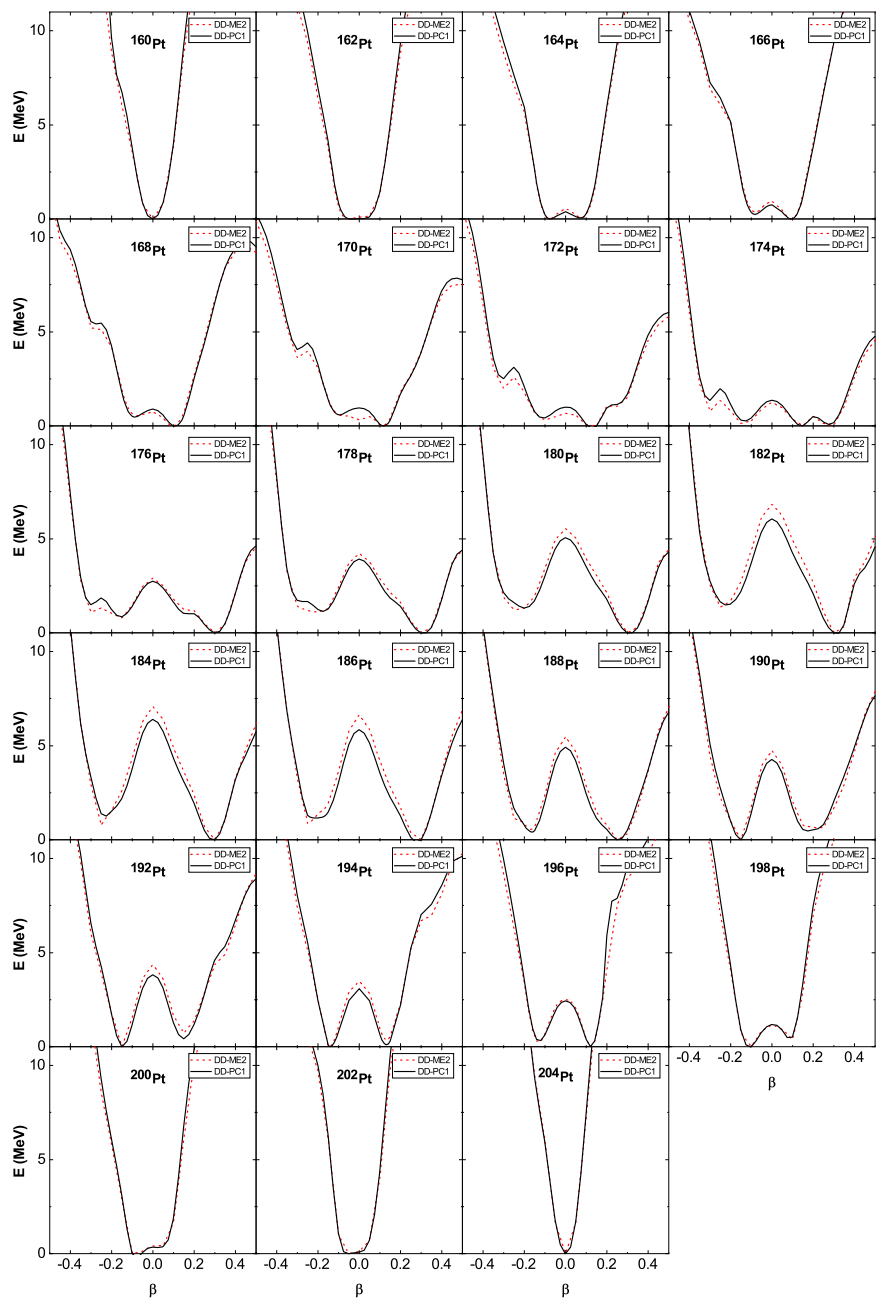}
	\caption{(Color online) The total energy curves for $^{160-204}$Pt as a function of the axial quadrupole deformation parameter $\beta_2$.}
	\label{23beta}
\end{figure}

The quadrupole deformation is also an important property for describing the
structure and shape of the nucleus. 
In Fig.~\ref{23beta}, we show for every Pt isotope (covering the mass
interval $160 \leqslant A \leqslant 204$) the energy curves along
the axial symmetry axis, as a function of the deformation
parameter, $\beta$, obtained within CDFT framework by using the density-dependent effective interactions DD-ME2 and DD-PC1.

As we can see from Fig.~\ref{23beta}, the interaction DD-PC1 provides potential
energy curves which are extremely similar to the ones obtained with DD-ME2. The deformations of the oblate and prolate minima are practically independent of the force. 

The lightest isotopes, $^{160-162}$Pt, exhibit spherical shape. The next isotope, $^{164}$Pt, starts to develop two shallow degenerate minima, oblate
and prolate, that correspond to a small value of $\beta$. The next isotope, $^{166}$Pt, starts to develop a more pronounced prolate minimum. The $^{168-186}$Pt isotopes show a similar structure, with a well-deformed prolate minimum, $\beta \approx 0.3$, and an oblate local minimum. 

A transition from prolate to oblate shapes occurs smoothly between $^{188}$Pt(prolate) and $^{190}$Pt(oblate). In $^{190-200}$Pt two minima appear, with the opposite situation
occurring in $^{168-186}$Pt. As the mass number increases, the two well-deformed minima gradually disappear and we get a flat potential energy curve at A = 202. At A = 204, we get a sharp single minimum, which confirms the spherical shape at the magic neutron number N = 126. 

These results are in good agreement with recent works \cite{Rodriguez2010,Garcia2014,Nomura2011} in terms of predicting $^{188}$Pt as critical. However, other calculations have different results that are not in agreement with ours, such as: Ref.~\refcite{Yao2013} which predicts that the shape transition in Pt isotopes within a beyond-mean-field approach with the Skyrme SLy6 occurs at A = 186 to 188 instead of A = 188 to 190 in our calculations. In the same line, 
constrained Hartree-Fock+ BCS calculations with the Skyrme forces Sk3, SGII, and SLy4 suggest a prolate to oblate shape transition at $^{182}$Pt~\cite{Boillos2015}. Furthermore, triaxial D1M-Gogny calculations predict a smooth shape transition at A = 184 to 186~\cite{Nomura2013}.

These differences between theoretical methods in predicting the exact location of the shape transition are due, firstly, to the difference between the models used and, secondly, to the fact that the shape transition is very sensitive to small details of the calculation, since the shape transition occurs exactly around the region where the energy of the competing shapes are practically degenerate.

\begin{figure}[!htb]
	\minipage{0.5\textwidth}
	\centering \includegraphics[scale=0.4]{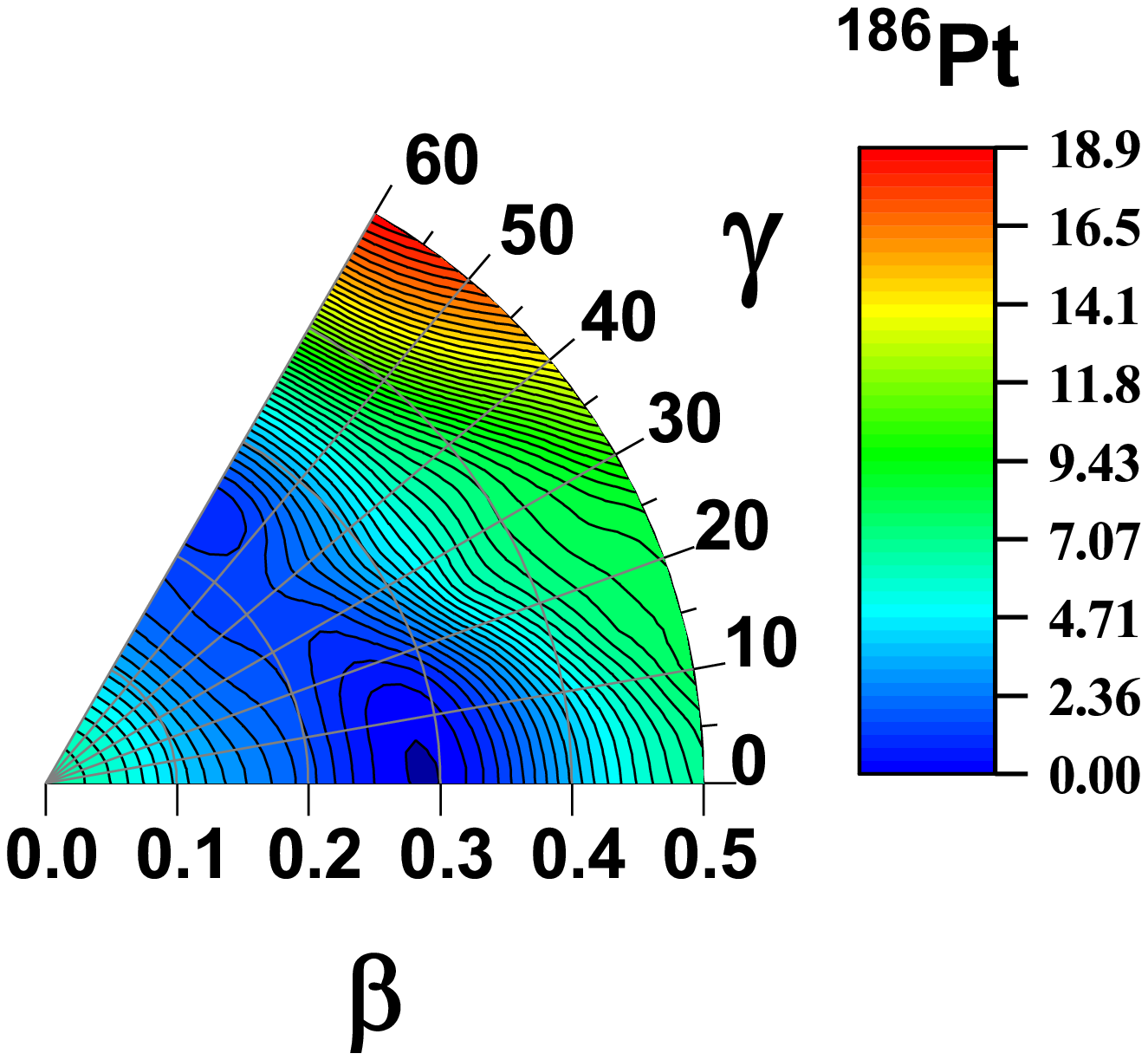}
	\endminipage\hfill
	\minipage{0.5\textwidth}
	\centering \includegraphics[scale=0.4]{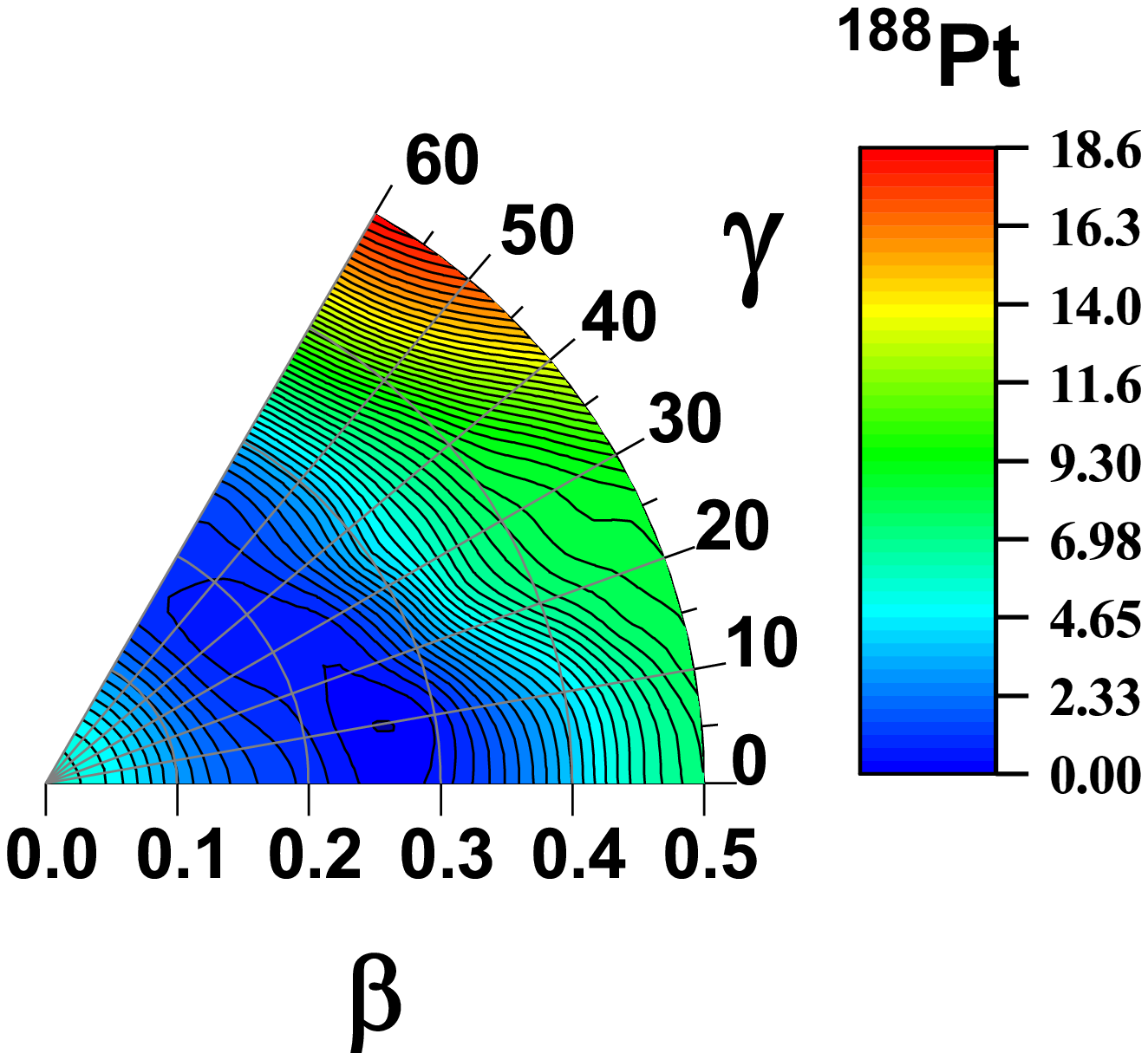}
	\endminipage\hfill
	\centering \includegraphics[scale=0.4]{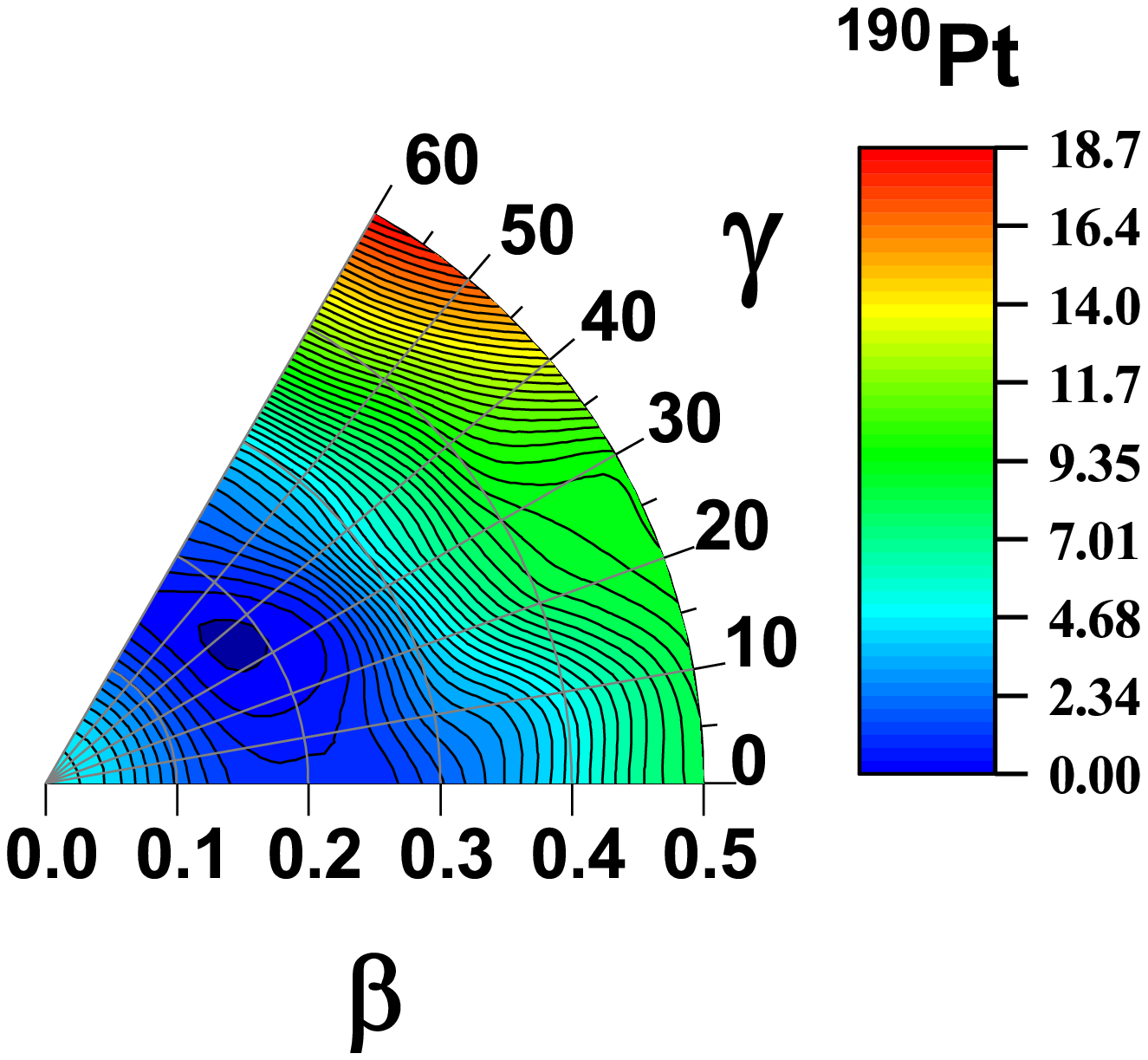}
	\caption{(Color online) Potential energy surfaces for  $^{186-190}$Pt in the ($\beta$, $\gamma$) plane, obtained from a CDFT calculations with the DD-ME2 parameter set. The color scale shown at the right has the unit of MeV, and scaled such that the ground state has a zero MeV energy.}
	\label{PESs}
\end{figure}

In Fig.~\ref{PESs} we display the triaxial contour plots of $^{186-190}$Pt
isotopes in the ($\beta$, $\gamma$) plane. To study the dependency on $\gamma$,  systematic constrained triaxial calculations have been done for mapping the quadrupole deformation space defined by $\beta_2$ and $\gamma$ using the effective interaction DD-ME2. 
The constrained calculations are performed by imposing constraints on both axial and triaxial mass quadrupole moments. The potential energy surface
(PES) study as a function of the quadrupole deformation parameter is performed by the method of quadratic constraint~\cite{Ring1980}  (see Ref.~\refcite{Abusara2017} for more details).
Energies are normalized with respect to the binding energy of the global minimum such that the ground state has a zero MeV energy.

From this figure, we can notice that the location of the ground state minimum moves from near prolate shape at $^{186}$Pt to near oblate shape at $^{190}$Pt. $^{188}$Pt is slightly triaxial with its global minimum at (0.25, 10\si{\degree}). Thus, the shape transition is smooth, and there are no sudden changes in the nuclear shape. These  results confirm those seen previously in Fig.~\ref{23beta} and are in full agreement with the results shown in Fig. 5 of Ref.~\cite{Garcia2014} obtained with Hartree-Fock-Bogoliubov based on Gogny-D1S interaction.

\subsection{Neutron, Proton and Charge radii}

\begin{figure}[ht]								   
	\centering \includegraphics[scale=0.55]{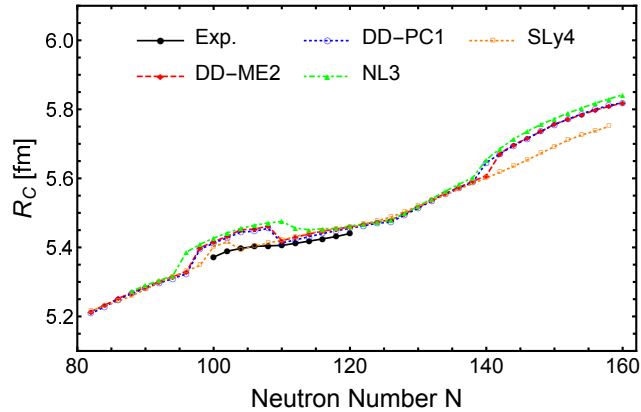}
	\caption{(Color online) The charge radii of Pt isotopes}
	\label{Rc}
\end{figure}

The charge radii calculated within the framework of the covariant density functional theory by using the functionals DD-ME2 and DD-PC1 are compared with the available experimental data~\cite{angeli2004} and with the predictions of RMF(NL3)~\cite{lalazissis1999} and HFB(SLy4). 
Good agreement between theory and experiment can be clearly seen in Fig.~\ref{Rc}, except in the region $100 \le N \le 108$ where a small difference between the models and experiment is seen. This discrepancy is mainly due to the deformation effect since the experimental $\beta_2$ values are extracted from experimental $B(E2)$$\uparrow$ values (the reduced probability of the transition from the ground state of the nucleus to the first excited $2^+$ state) by using the Bohr model, which is valid only in the case of well-deformed nuclei, as it has been explained in Ref.~\refcite{elbassem2019}.\\

\begin{figure}[!htb]
	\minipage{0.48\textwidth}
	\centering \includegraphics[scale=0.4]{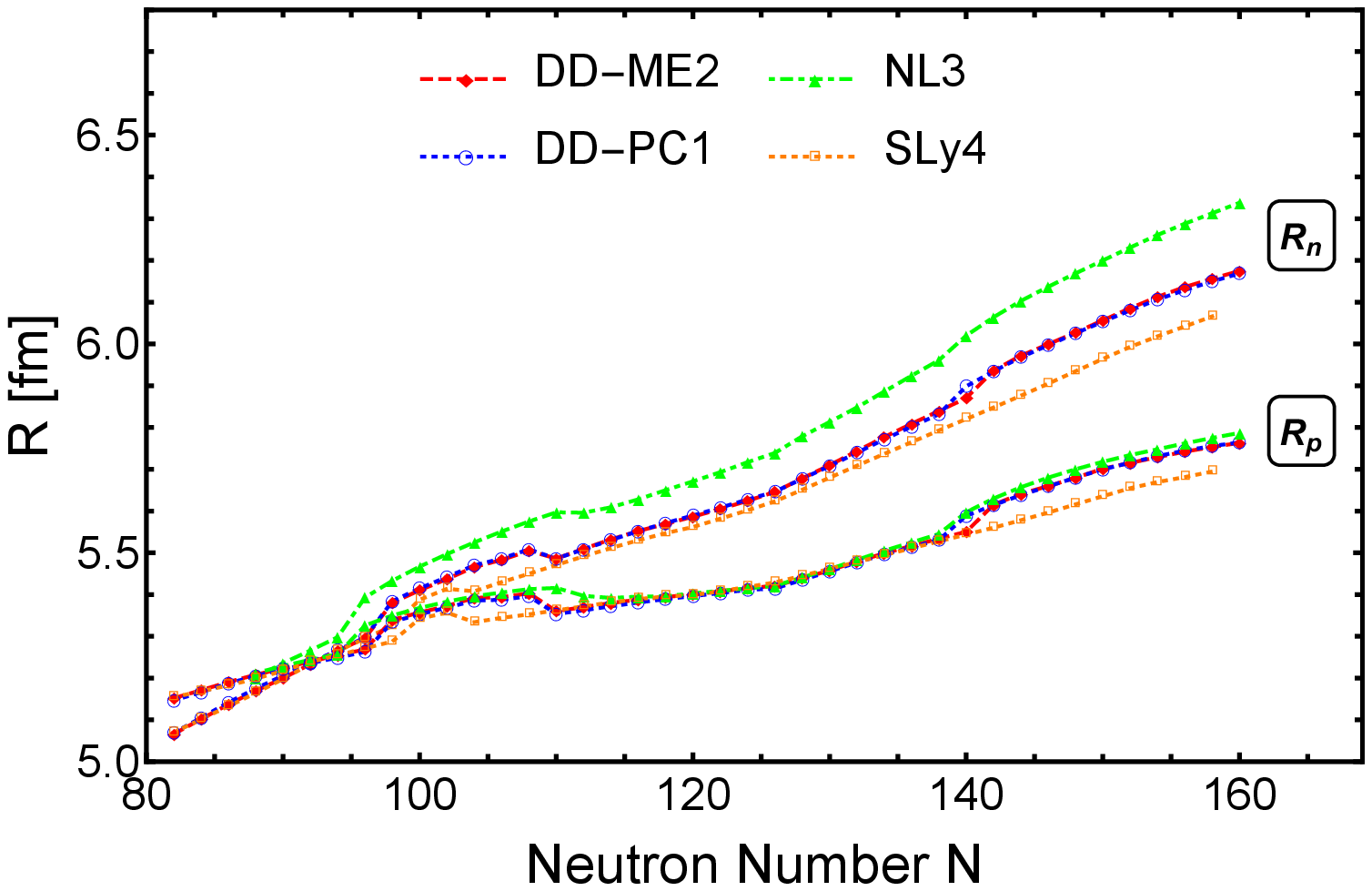}
	\endminipage\hfill
	\minipage{0.48\textwidth}
	\centering \includegraphics[scale=0.41]{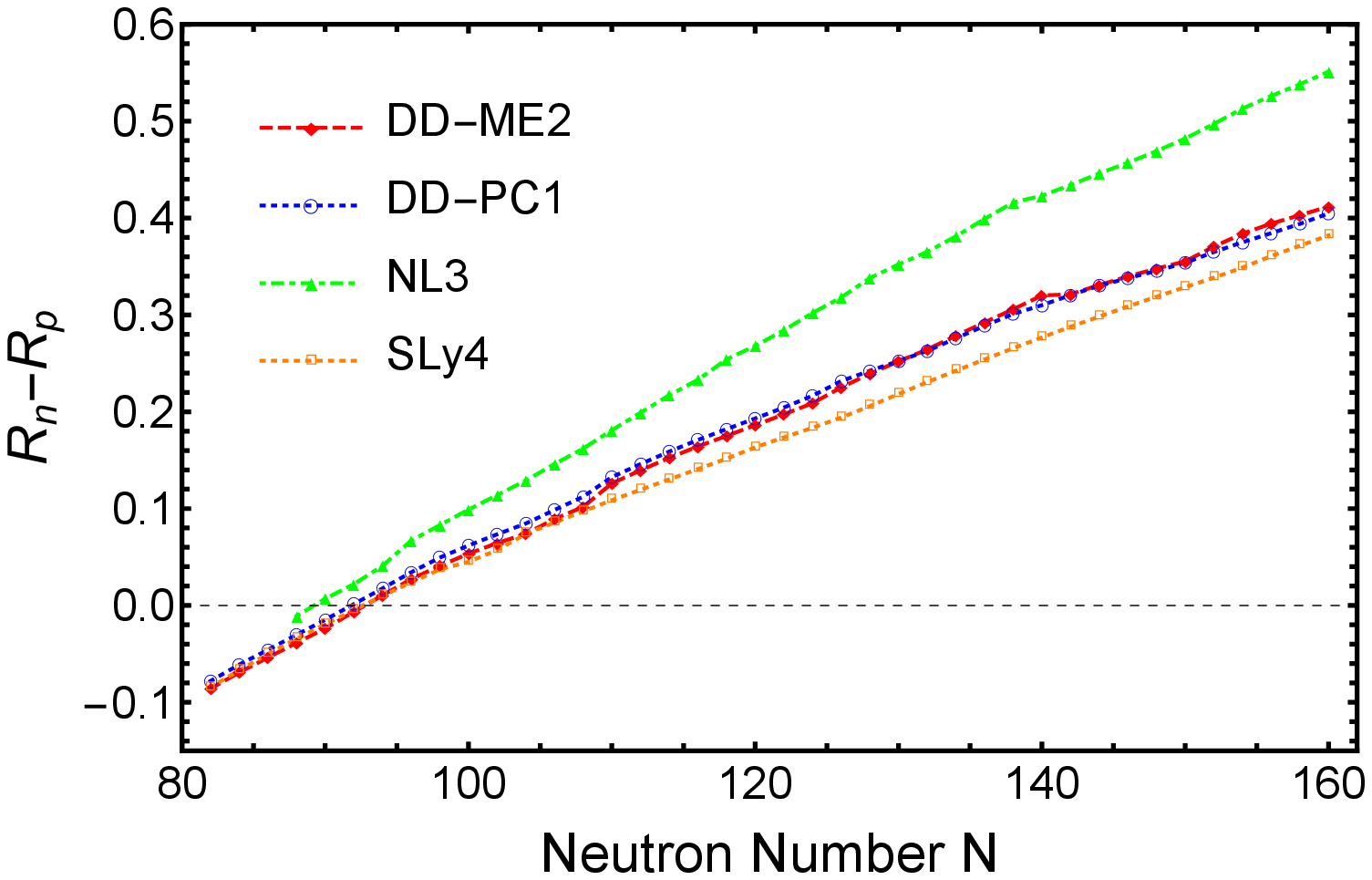}
	\endminipage\hfill
	\caption{(Color online) The neutron and proton radii of Pt isotopes (left panel) and the neutron skin thicknesses ($\Delta R$ = $R_n$ − $R_p$)(right panel).}
	\label{RnRp}
\end{figure}

We display in Fig. \ref{RnRp} the calculated neutron and proton radii, $R_n$ and $R_p$, of platinum isotopes obtained by using the two functionals DD-ME2 and DD-PC1. The results of HFB theory with SLy4 Skyrme force as well as those of RMF model with NL3 functional are also shown for comparison.

In the left panel of Fig.~\ref{RnRp}, we see that the neutron rms radii obtained by using the density-dependent
effective interactions DD-ME2 and DD-PC1 and by HFB(SLy4) are almost equal
throughout the isotopic chain, but the NL3 results are overestimated. The origin of this difference is the missing density-dependence in the iso-vector channel of NL3~\cite{lalazissis2005}. The proton rms radii obtained by all the formalisms, CDFT(DD-ME2 and DD-PC1), HFB(SLy4) and RMF(NL3) are almost similar.

We also notice from Fig.~\ref{RnRp} (right panel) that the difference between the rms radii of neutrons and protons ($\Delta R = R_n - R_p$) increases by increasing the neutron number in favor of developing a neutron skin. 
$\Delta R$ attains its maximum for $^{238}$Pt in which it reaches 0.41~fm in the case of DD-ME2 and DD-PC1 formalisms, 0.38~fm in HFB(SLy4) and 0.55~fm in RMF(NL3) calculations.

There are also some abnormalities in the charge and proton radii of Pt isotopes as we can see in Figs.~\ref{Rc} and~\ref{RnRp} (left panel), these little jumps in $R_c$ and $R_p$ in the two regions $98 \leqslant N \leqslant 108$ and $140 \leqslant N \leqslant 160$ are due to the deformation effect~\cite{elbassem2019}.\\

\section{Conclusion}
\label{Conclusion}
In the present work, we have studied the ground state properties of even-even
platinum isotopes, $^{160-238}$Pt, from the proton-rich side up to the neutron-rich one within the framework of the covariant density functional theory, by using two of the most recent functionals: The density-dependent point-coupling DD-PC1 and the density-dependent meson-exchange DD-ME2.
The bulk ground state properties are quite well reproduced in our calculations and are in good agreement with the experimental data.
A strong shell closure is clearly seen at N=126. The neutron skin in this study reaches 0.41 fm for $^{238}$Pt.  
The total energy curves for $^{160-204}$Pt obtained in this work suggest a smooth
prolate to oblate shape transition at $^{188}$Pt.\\

	
\end{document}